# "One Time Pad" Password Protection:
## Using T.E.C. Steganography and Secure Password Transmission Protocols

Givon Zirkind
B.Sc., Computer Science, Touro College; M.Sc. Computer Science, Fairleigh Dickinson University
givonzirkdin@gmail.com

## ABSTRACT

A while ago, I developed what I called an encryption method. The most favorable of reviews did not see a method but a collection of techniques. (The use of base Fibonacci being the most interesting.) Be that as it may, whatever process is being used, is described in the original paper, "Windtalking Computers" [ZIR01]. This paper is about the steganographic method described in ZIR01, the cryptanalysis efforts of that method and; a real world application of that method as an answer to the increasing problem of password file hacking. The premise is that the technique is a variant of one time pad, using a novel way to produce one time pad output for digital input. There is no record in the literature of such a method being used for encryption at all. This includes being used for one time pad encryption. Digital encryption generally treats the letters of the plaintext as a binary number and does some mathematical computation to produce ciphertext. The idea of <u>inserting bits</u> with a random generated key is new. Therefore (because a uniquely random generated key is used), the encryption is cryptanalytically unbreakable and/or computationally secure and/or information theoretic. An academic version was made. Challenges for decryption have not produced to-date a decryption. Advantages and disadvantages of the method are discussed.

Hackers are constantly penetrating networks and stealing password files. Which, once in possession of a password file, hackers individually or collectively with distributed processing over the Internet, decrypt the values of the hash passwords. Thereby gaining access to systems. This problem has become sufficiently significant for CAESAR (Competition for Authenticated Encryption: Security, Applicability, and Robustness) [CAE01] to make calls for papers for solutions. Herein is one proposed solution. While one time pad presents a problem being computationally intensive, for the relatively short length of passwords, the cost of computation may be cost effective for the security provided.

## Categories and Subject Descriptors
D.2.11 [**Software**]: Software Architectures – *Data abstraction*
E.0 [**Data**]: General
E.3 [**Data**]: Encryption
E.3 [**Data**]: Encryption – *Code Breaking*
E.m [**Data**]: Miscellaneous
F.2.0 [**Theory of Computation**]: Analysis Of Algorithms And Problem Complexity – *General*
F.2.1 [**Theory of Computation**]: Analysis Of Algorithms And Problem Complexity – Numerical Algorithms and Problems – *Number-theoretic computations*
F.2.2 [**Theory of Computation**]: Analysis Of Algorithms And Problem Complexity – Nonnumerical Algorithms and Problems – *Pattern matching*
F.2.m [**Theory of Computation**]: Miscellaneous
H.0 [**Information Systems**]: General
H.1.0 [**Information Systems**]: Models and Principles – General
H.1.1 [**Information Systems**]: Models and Principles – Systems and Information Theory – *Information Theory*

## 1. INTRODUCTION

Due to the increasing frequency of password files being captured and hacked over the Internet, CAESER has made a call for papers for suggested solutions to the problem. This paper proposes a solution to the problem. The solution involves a unique and novel variant of one time pad applied with a variation of a common session key issuing protocol. As the size of passwords are small, the cost of computation of unique non-repeating keys is reasonable.

## 2. T.E.C. ALGORITHM DEFINITION

T.E.C. is described as 2 separate methods / categories of encryptions or methodologies of encryption. One method is a form of steganography which uses bit shifting and the insertion of bits to mask the original byte stream. The supposition is, that the T.E.C. steganographic codec is analogous to one time pad.

Although not necessary for the T.E.C. steganographic method to be effective, a premise of some of the T.E.C. methods posits that using certain techniques in combination causes certain combinatorics that are relevant to either making brute force computationally unfeasible or cryptoanalytically unbreakable. As I will explain, these techniques are not necessary for the security of the stenagraphy. It is an added benefit to use one of the methods described of using an alternate binary representation other than base 2 [WIK01][WIK03], but not necessary for the T.E.C. steganographic method to be "prefectly secure". In and of itself, using an alternate binary base is merely obscurity security. Using an alternate binary representation other than base 2, increases the many possible binary representations necessarily considered for decryption and; will increase the number of bytes used to represent a character. [See [HUF01][SHA01] [ZIR01]. Base 2 is the most compact compressed binary representation.]

As ZIR01 posits, almost all encryption targets the letters themselves or their numeric value. T.E.C. on the other hand, targets the binary representation and does so separately from attempting to encrypt the written language. There is no encryption of the letters per se. The encryption is solely of the binary representation.

## 3. CRYPTANALYSIS

### 3.1 Fundamentals

Kerckhoff's principle [KER01] and T.E.C.

Conventional crytologic wisdom has it that whatever we know, our attackers know. The attackers know the algorithm. The attackers can get their hands on the ciphertext. But, they don't have the key. [KAH01] [KER01]

To make T.E.C. analogous to this cryptologic theory that the attackers know what we know:

- An article describing the encryption processes is published.
- The T.E.C. steganographic technique interjects additional <u>bits</u> into the <u>byte</u> stream to alter the <u>byte stream</u>.
- A random key is generated from a transcendental number that is used to choose how many bits to insert into each byte.
- An academic version is available. We can download the academic version and use it to generate specific ciphertext, given specific plaintext. Then, we can try to cryptanalyze the ciphertext.
- Other techniques may be used in combination with the bit insertion steganography. Ex. Using an alternate binary representation, a mathematical computation on the plaintext, a conventional key encryption of the plaintext, etc.

### 3.2 Cryptanalysis of the T.E.C. Steganographic Method

Cryptanalysis of the steganographic method as presented in ZIR001 is straightforward. There is an obvious weakness in the example presented in ZIR001. (So, I won't belabor explaining the weakness.) However, as an example of implementing bit shifting as an encryption method, the example is good. As reviewers have noted, if enough randomness is used with this technique, then it becomes the equivalent of one time pad. Albeit implemented in a very different way from conventional keying. The selection of unique random keys is all that is required to fulfill this requirement.

The novelty of T.E.C. is the bit shifting and bit insertion for steganography and; to use bit shifting insertion as a one time pad.

Also, as described in ZIR01,the steganography could be used in combination with the an alternate binary representation. This would complicate matters more. But, any one of those methods or combination of methods is irrelevant. The strength of T.E.C. steganography lies in the random non-repeating key.

Also, as the name transcendental encryption codec implies, transcendental numbers are used for random sequencing. Reviewers have noted that using a transcendental like Pi is not really random. Because, the sequence is well known. This is true. In response, as posited in ZIR01, the codec can use many unique transcendental numbers, for each user, or even each instance. Since, "an infinite number of transcendentals exist". The math to do this does exist (and could include the use of personal identifiers and timedate stamps so uniqueness is guaranteed). For a few examples of the number theory supporting this statement:

Set 1:

$\{ \aleph_1 |$

$\{ T' \ \forall_{T'} = f(T_{1-\infty}, x_{1-\infty}) \ \ x \neq T, x \neq -T, x \neq T^{-1} \} \}$

[The subsets of $\aleph_1$, where all the transcendental numbers are the function of any transcendental and any another number, if the other number is neither the transcendental itself, nor the additive nor multiplicative inverse of the primary transcendental number.]

or,

Set 2:

{ ℵ₁ |

{ T' ∀$_{T'}$ = f(T$_{1-\infty}$,x)  x ≠ T, x ≠ -T, x ≠ T$^{-1}$ } }

[The subsets of ℵ₁, where all the transcendental numbers are the function of any transcendental and another number, if the other number is neither the transcendental itself, nor additive nor multiplicative inverse of the primary transcendental number.]

or,

Set 3:

{ ℵ₁ |

{ T' ∀$_{T'}$ = f(T,x$_{1-\infty}$)  x ≠ T, x ≠ -T, x ≠ T$^{-1}$ } }

[The subsets of ℵ₁, where all the transcendental numbers are the function of a transcendental and some other number, if the other number is neither the transcendental itself, nor the additive nor multiplicative inverse of the primary transcendental number.]

[There are sufficient personal identifiers to use an 'x', that a unique combination is guaranteed. The addition of a time date stamp to any combination of personal identifiers would surely guarantee a unique number.]

Let us proceed, arguendo, the codec can do this—generate an infinite number of unique random numbers so each encryption will be unique.

The promoted advantage of the steganographic codec is that it can generate an unlimited supply of unique non-repeating sequences "on the fly". The weakness to this is that these sequences are generated by some kind of mathematical formula. If we know the input parameters to the equation, we can also calculate the sequence. So, in a certain sense, the mask or key is not random. [SCH01]

However, how the sequence was interjected is also unknown. So, even if the key is known, the position of the key (necessary for decryption) is not known. This would make for a large number of "tries" for a brute force decryption. A minimum of 4 tries / byte (2^2), and a maximum of 8 tries / byte (2^3), raised to the power of the number of characters in the message.

For a 200 character message (an average message) [FRI01], this would require a minimum of (2^2)^200 = 2^400 possible tries (decryptions) and; a maximum of (2^3)^200=2^600 possible tries (decrytions). Since brute fails when the number of tries reach 2^20 or 2^40 [SCH01], then T.E.C. is by the same logic computationally secure, for message of 200 characters minimum.

Using the minimum parameters, 4 tries per byte, 2^2; a password of a required length of 10 or 20 characters would need (2^2)*10 and (2^2)*20 to break with brute force. This renders T.E.C. steganographic encrypted passwords computationally secure.

If maximum parameters, 8 tries per byte, 2^3 are necessary; then a 5 character password encrypted with T.E.C. steganography would be sufficient to be resistant to brute force and be computationally secure. [Since (2^2)*10 = (2^3)*5.]

If an alternate binary representation is used, it will generate a longer ciphertext with more possible positions and increase the number of tries necessary for brute force decryption. Using a combinations of techniques is a basic theme of T.E.C. Something that must be considered while decrypting.

A protocol demanding a 10 character password is all that is necessary to secure a password file with T.E.C.

In addition, T.E.C. steganographic as a one time pad, is information theoretic producing several false positives.

There is a counter measure that can be built into the software that in addition to all other parameters, the user can enter, add, parameters to the formula. So, in addition to the values that must be entered into the formula, values that may be hacked or social engineered or guessed, the actual formula itself becomes as unique as a key in other encryption systems. (Of course, the parameters could also be social engineered.) The entire formula may only be known to the user alone. Of course, the weakness here is not in the system, but in spoofing (with a false data entry screen for example) to trap the formula parameters (the equivalent of the password in this system). [MIT01][MIT02]

It appears that the T.E.C. steganographic version has the same requirements as one time pad [SHA01]—a unique key equal to the length of the transmission. Then, T.E.C. steganograhpic is perfect security. One must have knowledge of the key to decrypt the encryption.

We can debate if transcendental numbers are normal (have a normal distribution of digits). No one really knows. [PRE01]

T.E.C. steganographic is truly innovative. It is an encryption that is an implementation of one time pad using steganography.

The implementation may will be math intensive. This may require a lot of computing power. A disadvantage. However, considering the length of a password—only 5 or 10 characters, and that these calculations will be done on the server side, the cost of computation vs. the security gained, may be cost effective.

T.E.C. steganography inherently increases the length of the output. This is counter almost all encryption schemes. However, this similarity to hashing [SCH01] would produce encrypted passwords in a ciphertext that would be longer than what the length of the plaintext password is. Obscurity security. But, better than storing passwords in the clear. However, unlike hashing, where the encrypted password ciphertext is one length for all plaintext passwords (as it is on Unix systems), T.E.C. produces variable length ciphertext. Since some other hashing schemes do the same, this variable length does not seem to be a drawback.

## 3.3 Dictionary Attacks

The success of a dictionary attack is more a function of the password selection protocol than the encryption of the passwords.

An example of a strong password selection protocol:

- ☺ Must use a capital and small letter.
- ☺ Must be a combination of letter-number-special character.
- ☺ Do not use names.
- ☺ Do not use common identifying numbers such as birth dates, home addresses, phone numbers or social security numbers.

Passwords selected with the above conditions have difficulty being guessed by dictionary attack.

Also, as T.E.C. uses the full 256 character set, not just the alpha-numeric subset Ex. 65-90 for A-Z. But even, control codes such as 0-13, the programming of attacks and analysis becomes more complicated. This has to be accounted for. This is obscurity security at most. However, it does raise the bar for the level of expertise required to attempt attacks and do cryptanalysis.

## 3.4 Capture of Passwords During Transmission

In an Internet application, if the user enters the password in plaintext and then sends the password in plaintext to the host,. The host then matches the submitted password with the password on file. This submission format has a security weakness that the plaintext password could be captured in transit. A significant problem that the CAESER call for papers wishes to address. However, the password is sent, encrypted or in the clear, if captured, the attacker can simply submit the captured password to gain access.

Because T.E.C. steganographic can generate an infinite number of unique keys, T.E.C., in combination with some standard key transmission protocol, presents a solution to this problem. A plugin can be used to generate or encrypt a message or key sent to the host. This encryption could be based on a timedate stamp or the encryption of another random number.

Then, there is the issue of the user having sufficient computing power to run the math intensive plugin. Again, as the calculations are for a small number of characters / numbers, this should be possible with acceptable performance.

A "man in the middle" approach might be used. The attacker could acquire the plugin; generate a spoofed plugin that accepts the password and parameters, then possibly does the math, if the attacker is able to reverse engineer the math used. This does present the problem of how the attacker would get the user to download and use the spoofed version of the plugin. There are several social engineering possibilities. [MIT01][MIT02] The use of certificates and signatures might be a counter measure to prevent the use of spoofed plugins.

Another security issue of capturing the password is keylogging. There is no software defense against hardware keylogging. Software keylogging has potential defenses. There are anti-virus vendors for this. T.E.C. steganographic does not prevent keylogging. (T.E.C. keyfiles protects against keylogging. But, that is not the subject of this paper. See the commercial version of T.E.C. for a discussion of keyfiles.)

## 3.5 Brute Force

Brute force fails against a theoretically cryptoanalytically unbreakable encryption method such as one time key encryption. This has been proven. [KAH01] [SCH01] The only question is, is T.E.C. steganography the equivalent of one time pad. Since the steganography is done with unique bit patterns representative of unique numbers, I maintain that T.E.C steganography is the equivalent of one time pad. Hence, T.E.C. steganography is cryptanalytically unbreakable, information theoretic and perfectly secure.

Storing passwords, adhering to an appropriate password protocol, along with T.E.C. steganography would provide a very strong encryption, much better than conventional hashing. This would secure the passwords in a password file at a much higher standard than is currently available, with hashing or otherwise.

Since every user will have unique personal identifiers, the transcendental created will be unique for each user. So, there is no possibility of extracting the individual passwords, lining them up and looking for clues to a key. Even assuming that would be possible for such short text.

## 3.6 In Sum

As I understand the problem, the major issue is that the password files are seized. Then, subsequently, either singly or with a distributed Internet network, either brute force attacked or cryptanalyzed for password extraction. T.E.C. presents one possible solution.

As for the capture of passwords during submission over the Internet, while this may be an issue, it is not the primary method of intrusion.

In sum, T.E.C. presents a solution to the major problem of stolen password files being decrypted.

## 4. SECURE PASSWORD SUBMISSION

### 4.1 Fundamentals

In transmission, the security of the password is more a function of protocol than the strength of encryption. The combination of the T.E.C. steganographic method and a protocol could provide much more protection than is currently available. The following protocol is one example.

The user opens an account. Submits a password; formula parameters and several identifiers. These identifiers do not have to be confidential information (such as birth date or government issued i.d. number). These identifiers can be answers to security questions the user generates. The identifiers correlate to a seed for the formulas for generation of a transcendental key.

The host chooses identifiers for each user. Any user or account information will do. The identifiers will correlate to a number and be used in the host's formula to generate a non-repeating sequence. Unlike the user's identifiers that the host must store, this information can be entered dynamically by the sysadmin.

The host independently, chooses a secret formula to encrypt all passwords. This formula is different from the individual users' formulas. This will require a restriction that no user can choose T-host for f(T,x).

The host does not reveal the host secret formula or user identifiers used as seeds for non-repeating number generation.

When the user wishes to login, the user requests a time date stamp from the host. The host selects one of the user's identifiers. Encrypts the time date stamp with the user's formula and one of the user's identifiers. The host temporarily retains this information (the time date stamp and user requesting to login in).

The host sends the encrypted time date stamp along with a token (a number) correlating to the user's identifier.

The user decrypts the time date stamp with the identifier and user formula.

The user encrypts the password with the time date stamp and an identifier as the seed for the user's formula. Then, the user sends the encrypted password, with a token for the identifier, to the host.

The host uses the user's formula, with the time date stamp and user identifier (correlating to the token received) as the seed to the formula to decrypt the submission. The host extracts the key.

The decrypts the user's key and compares to the key submitted. If the two match, login is completed.

### 4.2 Protocol Steps

1. INITIALIZATION: The user opens an account. Submits a password; formula parameters and several identifiers.
2. The host chooses identifiers for each user.
3. The host independently chooses a unique secret formula to encrypt the user's password.
4. The host does not reveal the host secret formula or user identifiers used as seeds for non-repeating number generation to encrypt the passwords. .
5. When the user wishes to login, the user requests a time date stamp from the host.
6. The host selects one of the user's identifiers. Encrypts the time date stamp with the user's formula and one of the user's identifiers.
7. The host temporarily retains this information (the time date stamp and user requesting to login in).
8. The host sends the encrypted time date stamp along with a token (a number) correlating to the user's identifier.
9. The user decrypts the time date stamp with the identifier and user formula.
10. The user encrypts the password with the time date stamp, using one of the user's personal identifiers as the seed for the user's formula.
11. The user sends the encrypted password, with a token for the identifier, to the host. (This step could be modified so that the user does NOT send the token. The host would have to use all the users identifiers and try several decryptions for a match,)
12. The host uses the user's formula, with the time date stamp and user identifier (correlating to the token received) as the seed to the formula to decrypt the submission. The host extracts the key.
13. The host uses the host's formula and proper user identifiers to extract the key from the password file.
14. The host compares the two keys.
15. The host erases the record of the request to login and the time date stamp issued.
16. If sufficient time passes without the user responding, the host erases the record of the request to login and the time data stamp issued.

### 4.3 Protocol Analysis

This protocol requires the user to have a plugin to do the calculations and splicing in order to submit a login.

This does not seem to be an issue. The expense of calculating the math should be negligible considering that a password will be between 8-20 characters (a standard for modern day systems).

The sysadmin must be trusted not to reveal the secret formula or choice of identifiers. But, this is not a major concern. The major concern is the theft of the password file and sufficient time to launch an attack.

This protocol will not prevent keylogging.

Whatever weakness there is in the transmission of the password, the password file itself is perfectly secure. Because the host's one time pad encryption of the passwords is done with a separate non-repeating sequence from the non-repeating sequence used in password transmission.

Using the time date stamp as part of the seed for the formula for the non-repeating sequence ensures uniqueness.

Having a copy of the login program and password is not enough to spoof the host and login. One will have to also know the parameters to the formula (which can be keylogged) and the identifiers (which can be keylogged). However, the identifiers are never explicitly transmitted. They are only referenced by token. This addresses the issue of sending passwords in the clear. This protocol does not send passwords in the clear.

### 4.4 The Formalized Protocol

H$f$ – Host Formula

$I_{1..In}$ – Identifiers 1...n

P – Password

$T_S$ – Time Date Stamp

$T_{1...n}$ – Tokens 1...n

Ta – time allotted for acceptable login

Tp – time passed for login procedure

U$f$ – User formula

Host Side Password Encryption:

1. $E(Hf(P,I_n))$

Login Request:

1. User requests login.
2. Host $E(Uf(T_S,I_n))$
3. Host retains $T_S$
4. Host $E(Uf(T_S)),I_n \rightarrow$ User
5. User $D(\ E(Uf(T_S))),I_n$
6. User $E(Uf(P,T_S,I_n))$
7. User $E(Uf(P,T_S,I_n)),I_n \rightarrow$ Host
8. Host $D(E(Uf(P,T_S,I_n)),I_n)$
9. Host $D(E(Hf(P,I_n)))$
10. Host compares $D(E(Hf(P,I_n)))==D(E(Uf(P,T_S,I_n))$
11. If comparison succeeds and Tp<=Ta; then login is made.
12. Host deletes $T_S$
13. Host deletes login request.

### 4.5 Other Protocols

This is just one possible protocol. Many similar conventional key exchange protocols could be adapted to use T.E.C.

### 5. Conclusion

In sum, T.E.C. steganograhpic presents a very strong encryption that could certainly be used for small messages, passwords, to provide very strong encryption.

Employing T.E.C. steganography for password encryption in a password would provide "perfect security" and be "cryptanalytically unbreakable".

**Author's Bio:** Givon Zirkind received his Bachelor's in Computer Science from Touro College and; his Master's in Computer Science from Fairleigh Dickinson University, both schools are located in the USA. His career has involved computer operations; software engineering; design and management of business applications with extensive database programming and management; Internet, web page design and implementation; e-commerce solutions, Google analytics, and Amazon reselling SEO; computer communications, data transfers and telecommunications; data conversion projects; reverse engineering of data and legacy software; being a published author and editor of a technical journal; teaching and; automated office support. His research work includes AFIS data compression and independent genetic database development and research. He may be reached at his email: givonzirkdin@gmail.com

### 6. ACKNOWLEDGMENTS

To my grandfather for all his support in all my endeavors.

Dr. Larry T. Ray, R.I.P., Ph.D. Mathematics/Computer Science, Stevens Institute of Technology (NJ), formerly professor of computer science, Fairleigh Dickinson University, for his mathematical evaluation and support in this project.

To Jack Lloyd and all those on his Cryptography List, randombits.net for all their input and comments. Travis H. Jeremy Stanley, and others.

Bruce Schneier for his books and his advice in his books and articles. Trying to break your own cipher will lead to creating a strong cipher. Probity also demands cryptanalysis of any cipher one develops.